\documentstyle[12pt,epsfig]{article}

\begin{document}
\newcommand{\be}{\begin{equation}}
\newcommand{\ee}{\end{equation}}        
\newcommand{\w}{wavelet}
\newcommand{\an}{analysis}


\begin{center}

{\bf  WAVELET--PATTERNS IN NUCLEUS--NUCLEUS COLLISIONS AT 158$A$ GeV }

\vspace{2mm}

I.M. Dremin, O.V. Ivanov, S.A. Kalinin, K.A. Kotelnikov, 
V.A. Nechitailo, N.G. Polukhina 
                    
\vspace{2mm}

{\it P.N.Lebedev Physical Institute, 117924 Moscow, Russia}

\end{center}

\begin{abstract}
We propose to use {\it multidimensional} \w\ \an\ for pattern recognition
in high multiplicity events and consider some examples.
Wavelet analysis reveals clustering phenomena in multiparticle production
processes. Results of event-by-event studies of some nucleus--nucleus 
collisions at 158$A$ GeV are reported. The clusterization topology is found 
to be different in different events. Typical patterns include jets and 
ring-like structures. They are demonstrated here for five high multiplicity 
events.

\end{abstract}
With the advent of RHIC and LHC, the events with up to 20000 charged particles 
produced will be available. The problem of presentation and analysis of these
high multiplicity events becomes rather non-trivial.
Each particle can be represented by a dot in the
3-dimensional phase space. Therefore, the distribution of these dots can be 
found for a single event. Different patterns formed by these dots in the phase 
space would correspond to different dynamics. 
At lower energies, it has been shown that the 
distribution averaged over all events possesses fractal properties \cite{dwdk}
which become however less pronounced for nucleus-nucleus collisions as 
compared to hadron interactions. It would be instructive to learn if the same
happens for individual events with multiplicity increasing.

When individual events are imaged visually, the human eye has a tendency
to observe different kinds of intricate patterns with dense clusters 
(spikes) and rarefied voids. However, the observed effects are often dominated 
by statistical fluctuations. The method of factorial moments was proposed 
\cite{bpes} to remove the statistical background in a global \an\ and it shows 
fractal properties even in event-by-event approach. The \w\ \an\ reveals the
{\it local} properties of any pattern in an individual event at various scales
and, moreover, avoids smooth polynomial trends and underlines the fluctuation
patterns. By choosing the strongest fluctuations, one hopes to get those
dynamical ones which exceed the statistical component. 

First attempts to use \w\ \an\ in multiparticle production go back to
P. Carruthers \cite{carr, lgca, gglc} who used \w s for diagonalisation of 
covariance matrices of some simplified cascade models. The proposals of
correlation studies in high multiplicity events with the help of \w s were 
promoted \cite{sboh, huan}, and also, in particular, for special correlations 
typical for the disoriented chiral condensate \cite{shth, nand}. The
\w\ transform of the pseudorapidity spectra of JACEE events was done in Ref. 
\cite{sboh}. In Ref. \cite{adk} \w s were first used to analyze two-dimensional 
patterns of fluctuations in the phase space of individual high multiplicity 
events of Pb-Pb interaction at energy 158 GeV per nucleon. Averaging over many 
events can hide the individual patterns. The event-by-event study provides 
further insight into the underlying dynamics of high multiplicity events where 
statistical fluctuations become less important.

Previously, some attempts \cite{addk, aaaa, cddh} to consider such events with 
different methods of treating the traditional projection and correlation 
measures revealed just that such substructures lead to spikes in
the angular (pseudorapidity) distribution and are somewhat jetty. Various
Monte Carlo simulations of the process were compared to the data and failed 
to describe this jettyness in its full strength. More detailed \an\ 
\cite{dlln, agab} of large statistics data on hadron-hadron interactions
(unfortunately, however, for rather low multiplicity) with
dense groups of particles separated showed some "anomaly" in the angular 
distribution of these groups.

We have analyzed nucleus-nucleus collisions at 158 GeV/nucleon and demonstrate 
here five high multiplicity events used for wavelet decomposition with the aim 
to study the patterns inherent to them.

The data was taken from the emulsion chamber experiment EMU-15 at CERN by the
group from Lebedev Physical Institute. The chamber contains thin lead target
of 300 $\mu $m thickness and 38 nuclear emulsion layers each of 50 $\mu $m
thickness deposited on the 25 $\mu $m mylar base \cite{ast}. The first layer 
is placed in direct contact with the target, the second layer is separated 
from it by 300 $\mu $m, the gaps between others vary but are not less than 
3 mm. The target and layers are placed perpendicular to the impinging beam
of Pb nuclei. This geometry provides good track resolution. The total thickness
of the nuclear emulsion in the chamber is about 0.07 cascade units. Such a small
size is very important for registration of central Pb-Pb collisions with many
hundreds of charged secondaries produced. The chamber was located in the 
transverse magnetic field of $B=2$T, and it allowed to analyze both angles and 
momenta (for curved enough tracks) of charged particles. For the present 
analysis, we use, however, the
information about the angles only and show here just five central Pb-Pb events
with the highest registered multiplicities from 1034 to 1221 charged particles.

The target diagrams of secondary particles distributions for these events are
shown in Fig. 1, where the radial distance from the center measures the polar
angle $\theta $, and the azimuthal angle $\phi $ is counted around the center.
One can sum over the azimuthal angle and plot the corresponding pseudorapidity
($\eta =-\log \tan \theta /2$) distributions shown in Fig. 2. The pronounced 
peaks ($\eta $-spikes) strongly exceeding expected 
statistical fluctuations are seen in individual events. This inhomogeneity in 
pseudorapidity can arise either due to a very strong jet i.e. a large group 
of particles close both in polar and azimuthal angles or due to a ring-like 
structure when several jets with smaller number of particles in each of them 
have similar polar angle but differ in their azimuthal angles\footnote{It can 
be called the tower substructure \cite{pesc} of the ring.} (in the 
limiting case, they can overlap and form a rather continuous ring\footnote{Then 
it is called the wall substructure \cite{pesc}.}). In all these cases spikes of
the pseudorapidity distribution ($\eta $-spikes) are observed. The possibility 
for one or several jets to have a predicted and similar polar angle was 
discussed in the framework of a hypothesis of Cherenkov gluons \cite{dr, dre2}.
According to it, each colliding nucleus is considered as a bunch of quarks
traversing the hadronic medium (another nucleus) and emitting gluons producing
final jets. Cherenkov gluons can be created if the nuclear refractivity index
exceeds 1. This excess is related to the positivity of the real part of the
forward elastic scattering amplitude for hadrons observed in experiment. Thus 
the necessary condition for this radiation type is fulfilled. The ring-like
structure can result from the gluon radiation at a short radiation length as
well. Depending on the mean number of gluons produced in a single event, the
resulting (the tower or the wall) pattern can differ.
It is a well defined position of the ring depending on the radiation 
length \cite{dr, dre2}  which is important for theoretical conclusions but not 
only the jettyness of the ring by itself. Moreover one can expect that the
statistical background is rather high if the probability of such radiation is 
low. 

As mentioned above, earlier \an\ \cite{aaaa, cddh} showed that even though 
the statistical fluctuations dominate in the azimuthal structure of ring-like
events and global characterization of all events is not conclusive, the real
sample shows larger jettyness than in the generated samples. For very dense and
isolated groups selected, their angular distribution \cite{dlln, agab} favoured
the theoretical predictions. At present we do not have high statistics of very
high multiplicity events to check the latest result and will concentrate here
on patterns in individual events. We rely on emulsion data because
inhomogeneity of acceptance in other detectors can not be corrected in the
event-by-event \an\ and thus results in ascribing the false patterns to 
individual events.

To reveal these patterns one should perform the 
two-dimensional local analysis. It is strongly desirable to get rid of such
drawback of the histogram method as fixed positions of bins that gives rise
to splitting of a jet into pieces contained in two or more bins. This chance
is provided by the wavelet transform of particle densities on the
two-dimensional plot. Wavelets choose automatically the size and shapes of 
bins (so-called Heisenberg windows; see, e.g., Ref. \cite{daub}) depending on 
particle densities at a given position. The multiresolution
analysis at different scales and in different regions is performed.

Previously, we presented the results of the one-dimensional analysis \cite{adk}
of the two-dimensional plot of one event numbered 19 in Figs.1 and 2 of that
paper. To proceed in this way, we subdivided the whole azimuthal 
region into 24 sectors (thus preserving the histogram drawback in this 
coordinate), analyzed the pseudorapidity distributions in each of them 
separately after integrating over azimuthal angles within a given sector, and 
connected them afterwards. Both jet and ring-like structures have been found
from the values of squared wavelet coefficients as seen from Figs. 
of Ref. \cite{adk}. At small scales, the wavelet analysis
reveals individual particles. At larger scales, the clusters or jets of 
particles are resolved. Finally, at ever larger scale we noticed the ring-like
structure around the center of the target diagram which penetrated from one 
azimuthal sector to another at almost constant 
value of the polar angle (pseudorapidity). For the event 19, this structure 
approximately corresponds to the peak in its histogram in Fig. 2 or to the ring
shown by dashed lines in Fig. 1 even though a part of it lies outside these 
regions (for more details see Ref. \cite{adk}). Moreover, it is not easy to 
notice in Fig. 1 any increase of density
within the ring just by eye because of the specific properties of the
$\theta - \phi $ plot where the density of particles decreases fast toward
the external region of large polar angles. 

Let us mention that other peculiar patterns were observed showing azimuthal 
asymmetry ($\phi $-spikes) in individual events which reminded the widely 
discussed (e.g.,see \cite{dman}) elliptic flow corresponding to the large
value of the second Fourier expansion coefficient or even three leaves
pattern with large third coefficient but we do not discuss them here.

The phenomenon of the particle density decrease at the target diagram 
outskirts is in charge of purely 
methodical sharp cut-off of all the pseudorapidity distributions at small 
rapidities seen in Fig. 2. The corresponding relative increase of the background 
in this region prevents precise particle detection, and therefore no 
experimental data is presented here. Let us stress that it is the only 
inhomogeneity in the acceptance of the emulsion chamber. It does not 
disturb the patterns at larger values of pseudorapidity. Since wavelet 
analysis is very sensitive to abrupt variations of functions, we have found in 
all five analyzed events large values of wavelet coefficients at $\eta \approx 
1.6-1.8$. Thus, wavelet analysis helps record the effects of the method of particle 
registration as well and therefore can be used for this purpose. In particular, 
we checked it by finding out the inhomogeneity of acceptance of some other 
detectors and showed that it changes the event patterns. In what follows we 
will not be interested in such effects leading to the ring-like (or other)
structures of purely methodical origin but concentrate on physics results
within the region of accurate registration of charged particles in the present
experiment.

In principle, the wavelet coefficients $W_{j_1,k_1,j_2,k_2}$ of the
two-dimensional function $f(\theta ,\phi )$ are found from the formula
\begin{equation}
W_{j_1,k_1,j_2,k_2}=\int f(\theta ,\phi )\psi (2^{-j_1}\theta -k_1;2^{-j_2}\phi
-k_2)d\theta d\phi .   \label{wav}
\end{equation}
Here $\theta _i, \phi _i$ are the polar and azimuthal angles of particles produced,
$f(\theta ,\phi )=\sum _i\delta (\theta -\theta _i)\delta (\phi -\phi _i)$
with a sum over all particles $i$ in a given event, $(k_1,k_2)$ denote the 
locations and $(j_1,j_2)$ the scales analyzed. The function $\psi $ is the 
analyzing wavelet. The higher the density fluctuations of particles in a given 
region, the larger are the corresponding wavelet coefficients.

In practice, we used discrete wavelets obtained from the tensor product of
two multiresolution analyses of standard one-dimensional Daubechies 8-tap
wavelets. Then the corresponding $ss, sd$ and $dd$ coefficients in the
two-dimensional matrix were calculated (see \cite{daub}). Within such an 
approach one should use the common scale $j_1=j_2=j$. Similar basis has been
used in Ref. \cite{goiv}.

As stressed above, the ring-like structure should be a collective effect
involving many particles and large scales. Therefore,
to get rid of the low-scale background due to individual particles and analyze
their clusterization properties, we have chosen the scales $j>5$ where both
single jets and those clustered in ring-like structures can be revealed as
is easily seen from Figs. of Ref. \cite{adk}. Therefore all coefficients 
with $j<6$ are put equal to 0. The wavelet coefficients for any $j$ 
from the interval $6\leq j\leq 10$ are now presented as functions of 
polar and azimuthal angles in a form of the two-dimensional landscape-like 
surface over this plane i.e. over the target diagram. Their inverse wavelet 
transform allows to get modified target diagrams of analyzed events with
large-scale structure left only. Higher fluctuations of particle density inside
large-scale formations and, consequently, larger wavelet coefficients
correspond to darker regions on this modified target
diagram shown in Fig. 3. Here we demonstrate two events (numbered 3 and 6) from
those five shown in Figs. 1 and 2. They clearly display both jet and ring-like 
structures which are different in different events. To diminish the role of
the methodical cut-off
at $\eta \approx 1.6-1.8$ we have used the requirement for the event to be 
approximately symmetrical in the region of low multiplicities and added 
to the unregistered region of small $\eta $ the tail appearing at large 
$\eta $ thus smoothing this cut-off. Therefore, the number of rings
corresponding to this methodical effects is lowered. However, the effect still
persists, and, to discard it, we shall consider the region of $\eta >1.8$ only.

Even though the statistics is very low, we attempted to plot the 
pseudorapidity distribution of the maxima of \w\ coefficients with the
hope to see if it reveals the peculiarities observed in high statistics but
low multiplicity hadron-hadron experiments \cite{dlln, agab}. In Fig. 4, the
number of highest maxima of wavelet coefficients exceeding the threshold value
$W_{j,k}>2\cdot 10^{-3}$ is plotted as a function of their pseudorapidity for
all five events considered. It is quite peculiar that positions of the maxima
are discrete. They are positioned quite symmetrically about the value 
$\eta \approx 2.9$ corresponding to $90^0$ in cms as it should be for two Pb 
nuclei colliding. Difference of heights is within the error bars.
More interesting, they do not fill in this central region but are rather
separated. Qualitatively, it coincides with findings in Refs. \cite{dlln, agab}.

For comparison, we generated 100 central Pb-Pb interactions with energy 
158 GeV/nucleon according to Fritiof model and the same number of events
according to the random model describing the inclusive rapidity distribution
shape. The fluctuations in these events are much smaller
than in experimental ones and do not show any ring-like structure.

Thus we conclude that even on the qualitative level there is the noticeable 
difference between experimental and simulated events with larger and somewhat 
ordered fluctuations in the former ones. We are not able to show all 
experimental and generated events because of lack of space but they just 
support our conclusion derived from those demonstrated above. With values of 
the wavelet coefficients at hand we could compare these events in a more
quantitative way presenting, e.g., the multiplicities in peaks, their number,
moments of the distributions etc. However, our aim here is to show the 
applicability of the two-dimensional analysis, the qualitative features and
differences leaving aside quantitative characteristics till higher statistics
of high multiplicity AA-events becomes available. In particular, the special 
automatic complex for emulsion processing with high space resolution in 
Lebedev Physical Institute (http://www.lebedev.ru/structure/pavicom/index.htm)
is coming into operation,
and we hope to enlarge strongly the statistics of analyzed central Pb-Pb 
collisions quite soon. The more complete \an\ of rings and statistics of their
angular positions will be available to compare with theoretical results.
Wavelets provide a powerful tool for event-by-event
analysis of fluctuation patterns in such collisions.

\newpage


\begin{center}
\vspace*{-1cm}
\epsfig{file=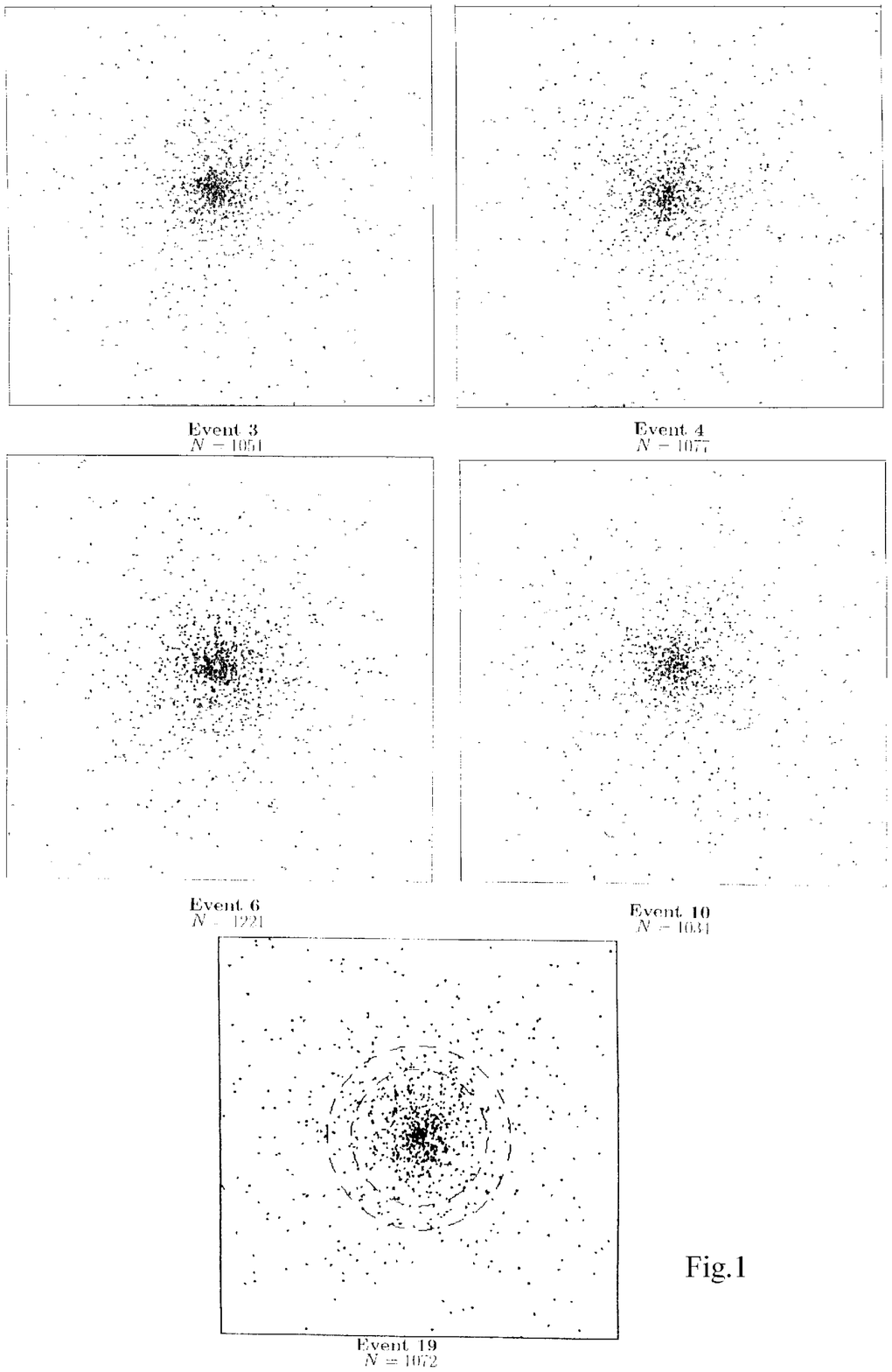,scale=0.65,clip=}
\end{center}

\vspace*{-1cm}
Fig.1 The target diagrams of five events of central Pb-Pb collisions 
         at energy 158 GeV/nucleon obtained by EMU-15 collaboration.

\begin{center}
\vspace*{-2.5cm}
\epsfig{file=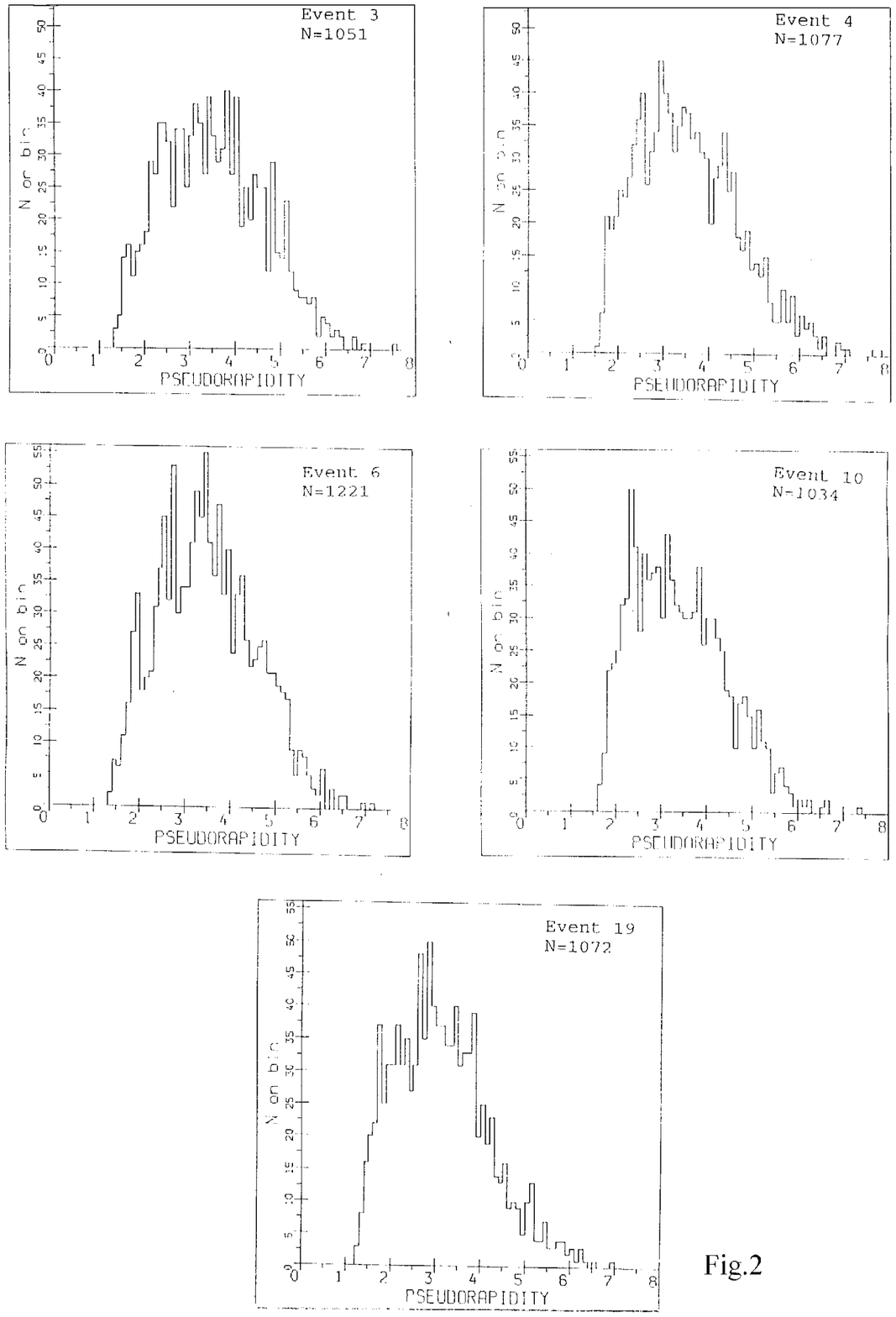,scale=0.7,clip=}
\end{center}
\vspace*{-1cm}
Fig.2 The pseudorapidity distributions of particles in five events shown
         in Fig.1.\\

\begin{center}
\fbox{\epsfig{file=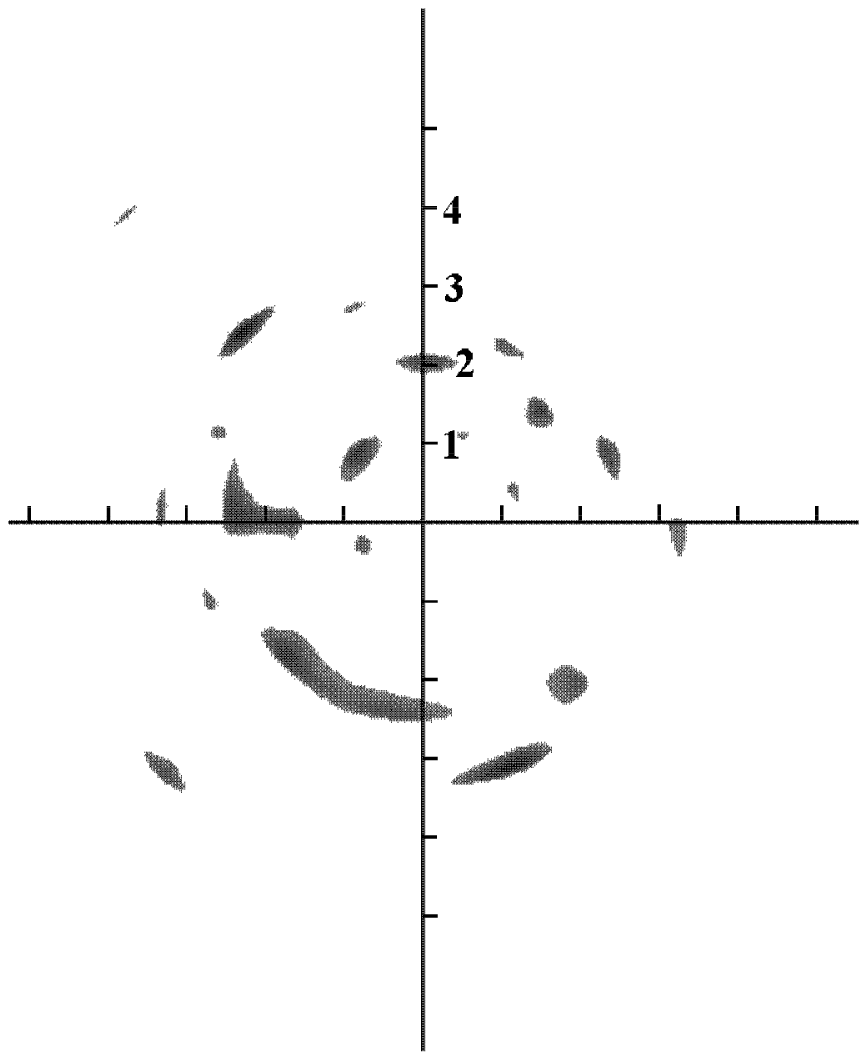,scale=0.47,clip=}}
\end{center}
\begin{center}
\fbox{\epsfig{file=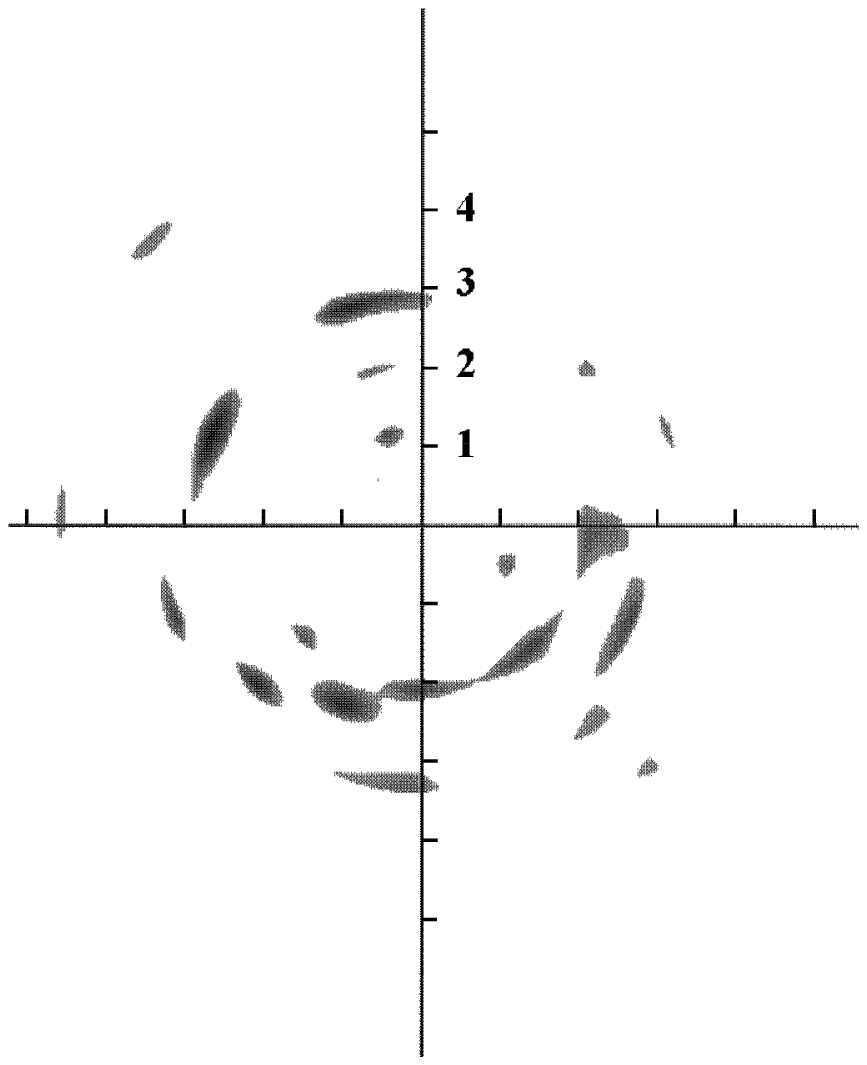,scale=0.47,clip=}}
\end{center}
Fig.3 The modified large-scale target diagrams of two events (3 and 6).
      Darker regions correspond to larger particle density fluctuations. \\

\begin{center}
\fbox{\epsfig{file=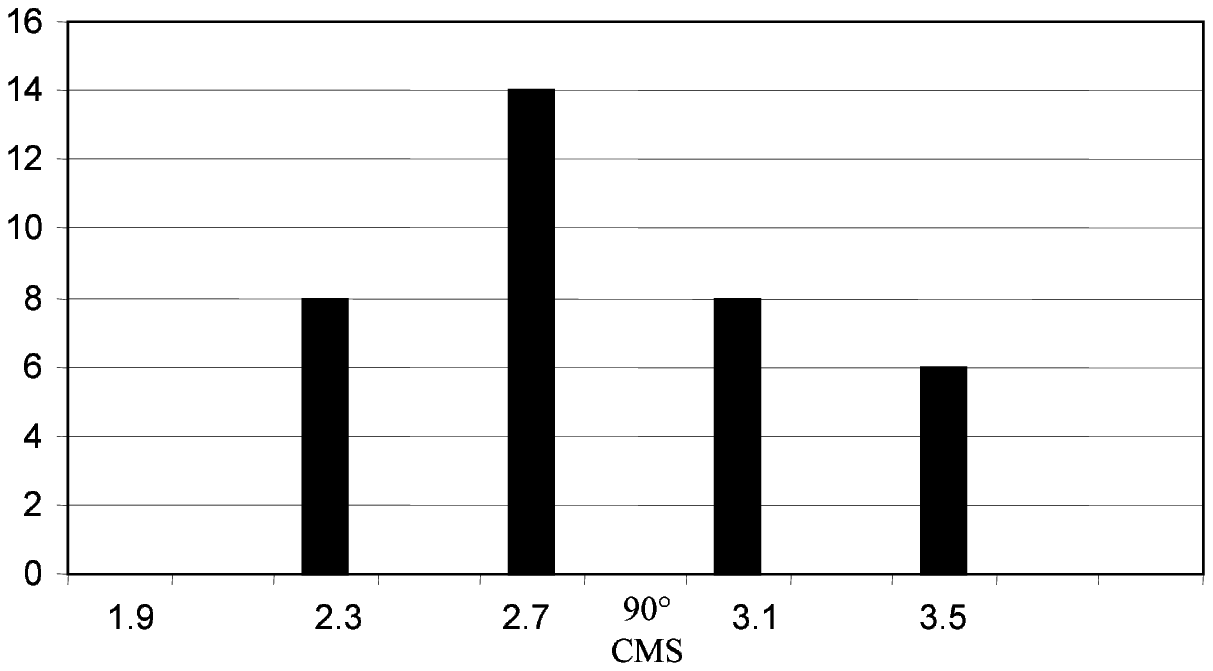,clip=}}
\end{center}
Fig.4 The pseudorapidity distribution of the maxima of \w\ coefficients.  \\
      The irregularity in the maxima positions, the empty voids between  \\
      them and absence of peaks at $\eta \approx 2.9$ are noticed. \\


\end{document}